\documentclass[slac_one]{revtex4}
\usepackage{graphicx}
\usepackage{fancyhdr}
\begin{document}

\noindent
\begin{minipage}[t]{.2\linewidth}
\leavevmode
 \hspace*{-.8cm}
\includegraphics{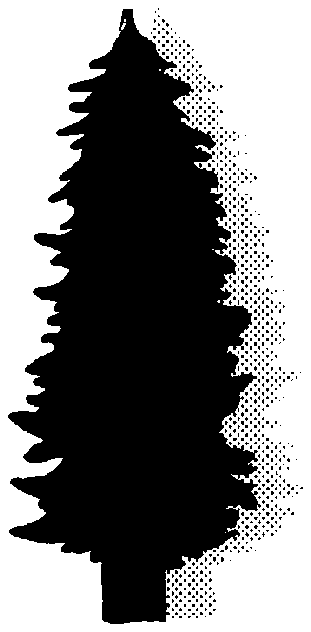}
\end{minipage} \hfill
\begin{minipage}[b]{.45\linewidth}
\rightline{SCIPP 05/05}
\rightline{July 2005}
\vspace{3cm}
\end{minipage}
\vskip1.5cm

\thispagestyle{empty}

\begin{center}
{\large\bf SELECTRON MASS RECONSTRUCTION AND THE RESOLUTION OF THE
LINEAR COLLIDER DETECTOR}\\[2pc]

SHARON GERBODE, HEATH HOLGUIN, TROY LAU, PAUL MOOSER, 
ADAM PEARLSTEIN, JOE ROSE, BRUCE SCHUMM \\

Santa Cruz Institute for Particle Physics \\
University of California, Santa Cruz
\end{center}
\vskip1cm

We have used ISAJET and the JAS LCD fast simulation to explore
the precision of Snowmass Point
SPS1a selectron mass reconstruction for the Silicon Detector
concept. Simulating collisions at
$E_{cm} = 1$ TeV, we have found that most of the information
constraining the selectron mass is carried in the forward
($|\cos \theta| \ge 0.8$) region. We have also found that,
for a beam energy spread of 1\% (conventional RF design),
detector resolution limitations compromise the
selectron mass reconstruction only in the forward region. However,
for a beam energy spread of less than 0.2\% (superconducting
RF design), the detector resolution compromises the selectron
mass reconstruction over the full angular region.

\vfill
\begin{center}
{\small Talk presented at the 2005 International Linear Collider
Workshop, Stanford University, Stanford, California, March 18-22, 2005}
\end{center}
\vfill
\clearpage
\setcounter{page}{1}

\title{{\small{2005 International Linear Collider Workshop -- 
Stanford, U.S.A.}}\\ 
\vspace{12pt}
Selectron Mass Reconstruction and the Resolution of the
Linear Collider Detector} 

%

\author{Sharon Gerbode, Heath Holguin,
Troy Lau, Paul Mooser, Adam Pearlstein, Joe Rose, Bruce Schumm}
\affiliation{SCIPP, University of California, Santa Cruz, CA 95064, USA}

\begin{abstract}
We have used ISAJET and the JAS LCD fast simulation to explore
the precision of Snowmass Point
SPS1a selectron mass reconstruction for the Silicon Detector
concept. Simulating collisions at 
$E_{cm} = 1$ TeV, we have found that most of the information
constraining the selectron mass is carried in the forward 
($|\cos \theta| \ge 0.8$) region. We have also found that,
for a beam energy spread of 1\% (conventional RF design),
detector resolution limitations compromise the
selectron mass reconstruction only in the forward region. However,
for a beam energy spread of less than 0.2\% (superconducting
RF design), the detector resolution compromises the selectron
mass reconstruction over the full angular region.

\end{abstract}

\maketitle

\thispagestyle{fancy}


\section{INTRODUCTION} 
For some time, the Santa Cruz Linear Collider R\&D group has been
interested in exploring the requirements that measuring forward
selectron production places on the Linear Collider Detector.
With right-handed selectron and lightest neutralino masses of 
143 and 95 GeV, respectively, Snowmass Point SPS1a produces
a substantial forward ($|\cos \theta| \ge 0.8|$) population of
electrons from selectron decay. Over the past two years, the
Santa Cruz group has developed techniques to isolate the
forward selectron-decay signal from Standard-Model backgrounds,
as well as techniques to extract the selectron mass from the
observed selectron-decay electron spectrum. This paper reports
on the results of those studies.

\section{SIGNAL SELECTION}

Backgrounds to the selectron-decay electron signal come from
two primary sources: four-electron production for which two of
the final-state fermions are close in angle to the beam trajectory,
and thus escape detection, and the $Z$-boson fusion process
$e^+ e^- \rightarrow e^+ e^- \nu \bar{\nu}$. During the course
of searches for supersymmetry at LEP, several cuts were developed
to isolate potential SUSY signals from Standard Model backgrounds.

\begin{itemize}

\item{ {\bf Fiducial Region Cut:} Exactly one final-state positron 
and one final-state electron pair must be detected in the fiducial
region. For this study, the fiducial region lies above 5 GeV in
momentum, and within 
$|\cos \theta| \le 0.8$ for central-region studies, or
within $|\cos \theta| \le 0.994$ for studies including the 
forward region.}

\item{ {\bf Tagging Cut:} 
No observable electron or positron in the low-angle `tagging
calorimetry' (with assumed coverage of 20mrad 
$< \theta <$ 110mrad)}.

\item{ {\bf Transverse Momentum Cut:} Cuts events for which
the vector sum of the observed
$e^+e^-$ pair transverse momenta is less than 
500 Gev $\times \sin$(20 mrad). }

\end{itemize}

For a signal region of $|\cos \theta| \le 0.8$,
these cuts completely
eliminate the four-electron background up to diagrams with additional
radiation, and sufficiently select against the $ee\nu\nu$ process.
However, as one attempts to reconstruct the selectron
decay signal beyond $|\cos \theta| = 0.8$, both the radiative
four-electron background and the $ee\nu\nu$ backgrounds approach
poles in their respective production cross sections, and begin
to dominate the selectron signal. To suppress these backgrounds
to a degree appropriate for use of the forward selectron decay
signal, two additional cuts were applied.

\begin{itemize}

\item{ {\bf Photon Cut:} Remove event if there is an observed photon
of 20 GeV or greater in either the fiducial or tagging region.}

\item{ {\bf High-Momentum Cut:} Remove even if the vector sum of
 the $e^+e^-$ pair momenta has a magnitude of greater than 225 GeV.}

\end{itemize}

With the application of these cuts to events produced 
at $E_{cm} = 1$ TeV with 80\%
right-handed-polarized electron beam and unpolarized
positrons, it was found that Standard Model processes 
provided a negligible background to the right-handed
selectron signal. In addition, the application of these
selection criteria did not significantly alter the shape
of the selectron-decay electron spectrum in either the
central or forward region. Nevertheless, 
in the analysis that follows all of the selection cuts
will be applied to the selectron signal; however, Standard
Model backgrounds will be ignored from this point on.
Finally, running with an 80\% right-polarized electron
beam, electrons from the decay of the more massive
left-handed selectron provide a diffuse background to
the right-handed selectron decay electron spectrum. 
This contribution was also ignored in the right-handed
selectron analysis.

\section{SIMULATION OF THE SELECTRON SIGNAL}

The selectron signal was generated at $E_{cm} = 1$ TeV using the 
Snowmass point SPS1a~\cite{sps} parameters
$m_0 = 100$ GeV, $m_{1/2} = 250$ GeV, $A_0 = -100$, 
$\tan \beta = 10$, and sgn($\mu$) = $+$. The SPS1a specifications
were implemented within the ISAJET~\cite{isajet} package, and
included the effects of initial state radiation, beamstrahlung
($Y = 0.29$), and three different values of the fractional beam
energy spread (assuming a gaussian distribution): 1.0\%, 
0.16\%, and 0.0\%. 

For this point in SUSY parameter space, 
the endpoint of the
selectron-decay electron energy distribution lies
above 270 GeV, making the precise measurement of
the endpoint energy by the tracking system
somewhat challenging. Nonetheless, the projected
$\sim {1 \over 2}\%$ resolution of the tracking system
at this energy is somewhat better than that expected
for the electromagnetic calorimetry. 

The 
$\cos \theta$ distribution of final-state electrons and 
positrons from the decay of right-handed selectrons is
peaked towards $|\cos \theta| = 1.0$, due to the sizeable
t-channel contribution admitted by the relatively light
selectron and neutralino. Approximately half of the signal
lies in the forward region, beyond 
$|\cos \theta| = 0.8$.

\section{DETERMINATION OF THE RIGHT-HANDED SELECTRON MASS}

To focus on the accuracy of the selectron mass reconstruction,
it was assumed for now that the lightest neutralino mass
was known precisely from other measurements. The Santa Cruz
group plans to relax this assumption in further studies.

The selectron mass was determined by finding the best fit
of the selectron energy spectrum to that of a series of
`template' distributions generated with 
slightly varying right-handed selectron masses, but
with all other aspects of the signal generation,
including beam energy spread and detector smearing effects,
the same as for the data simulation. Templates were produced
for 15 right-handed selectron masses in a range of approximately
$\pm 1$ GeV about the nominal SPS1a mass of 143.11 GeV.

A total of 120 independent data sets, each corresponding to
an integrated luminosity of 115 fb$^{-1}$, were generated
at the nominal SPS1a mass. For each of these data sets,
a $\chi^2$ was formed against each of the templates points,
according to
\begin{equation}\label{eq:chisq}
\chi^2 = \sum_{i}{(w * n_i - w * m_i)^2 \over
            (w * n_i - m_i)^2}
\end{equation}
where the sum is over energy bins in the selectron-decay
electron spectrum, $n_i$ and $m_i$ are the data and
template bin contents, respectively, and
\begin{equation}\label{eq:wieght}
w = {\sum_{i} n_i \over \sum_{i} m_i}
\end{equation}
is the relative sample-size weighting factor. To minimize
the statistical contributions of the template files, they
were generated with an integrated luminosity of approximately
1000 fb$^{-1}$ each.

For each data set, the $\chi^2$ contour was fit with a 
quartic polynomial, which was then 
minimized to find the best-fit selectron mass.
This procedure was repeated for the 120 data sets,
and the mean and root-mean-square deviation was
calculated from the distribution of best-fit
right-handed selectron masses.

In forming the $\chi^2$ for each of the 120
independent data sets, it was found that the
inclusion of energy-spectrum bins not sufficiently
close to the upper and lower endpoints to provide
useful information on the endpoint location 
introduced unacceptable scatter in the contour.
Thus, for each scenario that was studied 
(beam energy spread, detector model, polar
angle reach), only bins near the upper and
lower endpoint were used in forming the 
$\chi^2$. The exact energy ranges used in the
$\chi^2$ calculation depended on the scenario
under study,
and was determined in a data-driven manner by
examining the spectrum in the region of the
endpoints in the generated data samples.
Table I shows the regions used in the forming
the $\chi^2$ for the different scenarios.

\subsection{Scenarios Explored}

To explore the effects of detector resolution, 
data sets and mass templates were generated 
assuming perfect detector resolution as well
as that of the SDMAR01 detector. Detector resolution
effects were incorporated via gaussian smearing
based on error matrices from 
Billior-based tracking error calculations
provided by the LCDTRK program~\cite{lcdtrk}.
Two different ranges in $|\cos \theta|$ were
explored: between 0.0 and 0.8 and between 0.0
and 0.994. Finally, three different values
of the beam energy spread were explored:
$\pm 1.0\%$ (to allow benchmark comparisons to
previous studies~\cite{uriel,yang}), $\pm 0.16\%$ 
(approximately that expected for the 
superconducting RF design), and, for comparison,
0.0\%.

\begin{table}[t]
\begin{center}
\caption{Endpoint Fit Regions}
\begin{tabular}{|l|c|c|}
\hline \textbf{Scenario(s)} & Lower Endpoint $\chi^2$ Range
 & Upper Endpoint $\chi^2$ Range 
\\
\hline Perfect detector resolution and/or no energy spread & 5.2--6.4 GeV & 
269.2--273.2 GeV \\
\hline realistic resolution; 0.16\% energy spread & 
5.2--6.4 GeV & 267.8--274.6 GeV  \\
\hline realistic resolution; 1.0\% energy spread &
5.2--6.4 GeV & 267.2--275.2 GeV  \\
\hline
\end{tabular}
\label{l2ea4-t1}
\end{center}
\end{table}

\section{RESULTS}

Figure 1 shows the root-mean-square deviation observed for the 
120 independent data sets for the different scenarios that 
were explored in the study (in all cases, the mean value of the 
fit mass was reasonably
consistent with the input vale of the right-handed selectron
mass of 143.1 GeV).

\begin{figure*}[t]
\centering
\scalebox{0.7}{\includegraphics{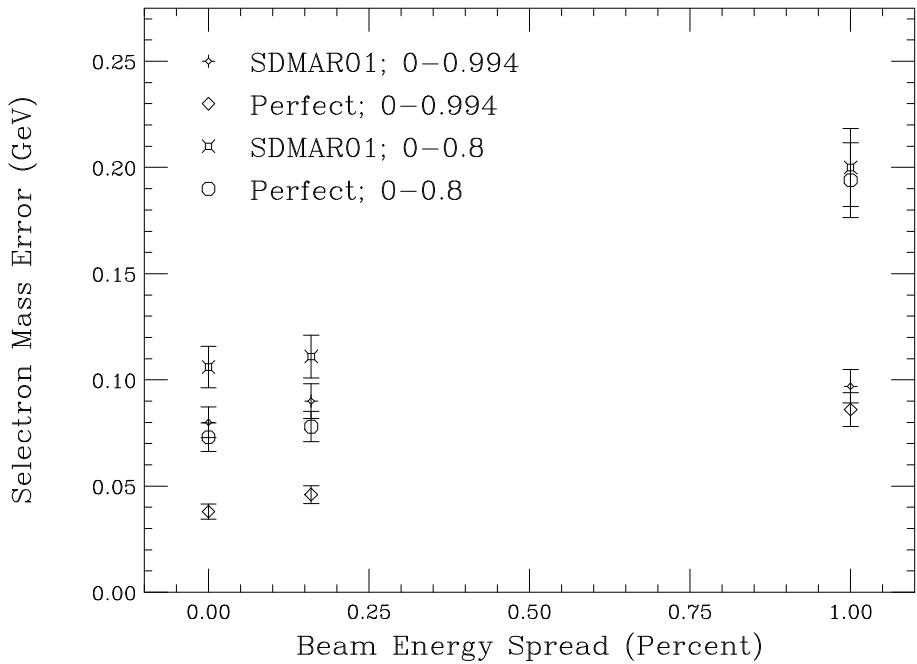}}
\caption{Sensitivity to the SPS1a right-handed selectron mass 
for 115 fb$^{-1}$ of Linear Collider data at $E_{cm} = 1000$ GeV,
as a function of beam energy spread and polar angle coverage.} \label{JACpic2-f1}
\end{figure*}

For a perfect detector (no measurement 
uncertainty), the inclusion of the forward region provides 
substantial improvement in the selectron mass measurement,
independent of beam energy spread. In fact, the improvement
is better than one would expect from the factor-of-two 
increase in the number of signal electrons that is
associated with the inclusion of the forward region.
A study of the
two-dimensional frequency distribution of signal electrons
as a function of energy and the cosine of the polar angle 
(Figure 2) reveals that the
spectrum has a greater contribution at higher energy, where
the sensitivity to the selectron mass is greatest, at high
values of $\cos \theta$. Thus, in this case, most of the 
information on the selectron mass (and on slepton masses
in general) is in the forward region.

\begin{figure*}[t]
\centering
\scalebox{0.7}{\includegraphics[angle=270.5]{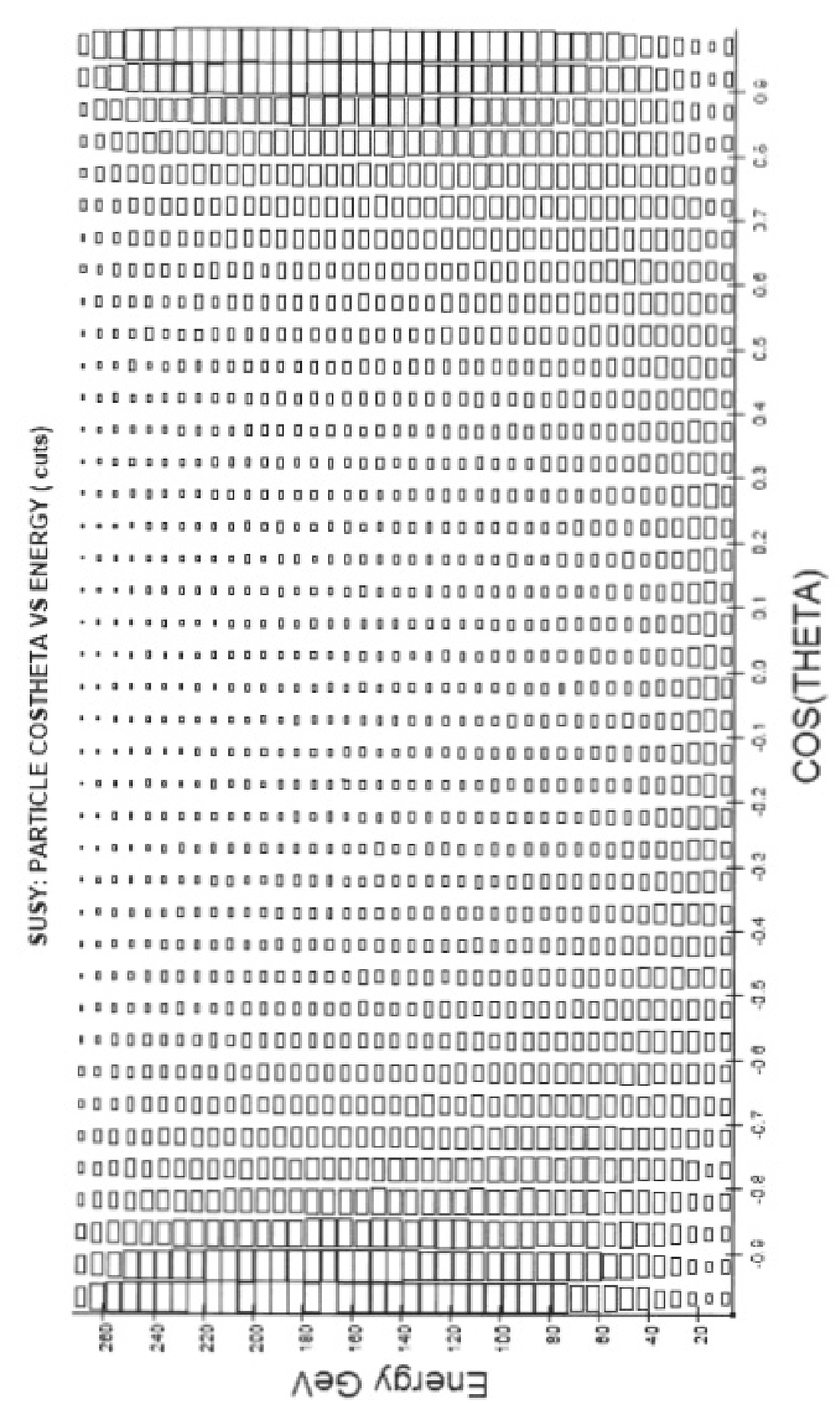}}
\caption{Frequency distribution of electrons from right-handed
selectron decay as a function of energy and polar angle.} \label{JACpic2-f2}
\end{figure*}

For large beam energy spread, it is seen that the 
sensitivity to the selectron mass has little dependence
on the detector resolution. For this case, the SiD design
seems to be adequate to take advantage of the physics
capabilities of the accelerator. However, for smaller
beam energy spread, this is not the case. For the energy
spread anticipated for the selected machine design, 
substantial improvement in the selectron mass determination
can be achieved by improving the detector resolution,
particularly in the forward region.

\section{CONCLUSIONS}

To examine the issue of measuring selectron masses in the forward
region of the Linear Collider Detector, we have simulated 
SPS1a selectron production at $E_{cm} = 1$ TeV. By developing
two new selection criteria, we have demonstrated that the
selectron signal can be separated from Standard Model 
backgrounds through the entire forward tracking region
$|\cos \theta| < 0.994$. 

Due to the stiffening of the selectron-decay electron spectrum
at higher values of $|\cos \theta|$, for a light selectron
most of the information on selectron mass (and slepton mass 
in general) comes from the forward region.
For large beam energy spread ($\pm 1.0\%$), we find that the
SiD detector design is adequate to exploit the potential of
the Linear Collider; however, for the expected beam energy
spread of the chosen cold RF technology, gains in sensitivity
to the selectron mass can be made by further improvements in
momentum resolution, in both the central and forward regions.

\begin{acknowledgments}

Although they appear as authors on this paper, one of us (Schumm)
would like to acknowledge the dedication and creativity of the
other authors, who performed these studies as undergraduate
physics majors at the University of California at Santa Cruz.

\end{acknowledgments}

\end{document}